\documentstyle[12pt,aps,floats]{revtex}
\begin{document}
\baselineskip=24pt
\input psfig
\baselineskip=18pt

\title{How the Sun Shines\footnote{Published in the Nobel e-Museum, 
http://www.nobel.se/physics/articles, June, 2000.} }
\author{John N. Bahcall}

\maketitle

\bigskip
What makes the sun shine? How does the sun produce the vast amount of
energy necessary to support life on earth?  These questions challenged
scientists for a hundred and fifty years, beginning in the middle of
the nineteenth century. Theoretical physicists battled geologists and
evolutionary biologists in a heated controversy over who had the
correct answer.

Why was there so much fuss about this scientific puzzle? 
The nineteenth-century astronomer
John Herschel described eloquently the fundamental role of sunshine in all
of human life in his 1833 {\it Treatise on Astronomy}:

\begin{quote}
The sun's rays are the ultimate source of almost every motion which 
takes place on the surface of the earth.  By its heat are produced all 
winds,...By their vivifying action vegetables are elaborated from 
inorganic matter, and become, in their turn, the support of animals and 
of man, and the sources of those great deposits of dynamical efficiency 
which are laid up for human use in our coal strata.
\end{quote}

\begin{figure}[!hb]
\centerline{\psfig{figure=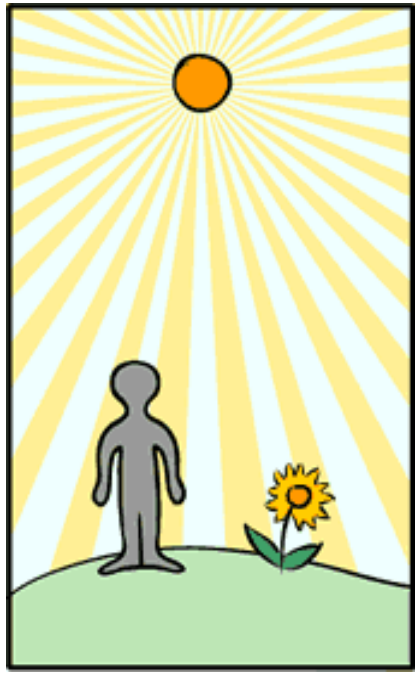,height=3.5in}}
\tightenlines
\caption[]{\footnotesize Sunshine makes  life possible on earth.
\label{fig:sungiveslife}}
\end{figure}

In this essay, we shall review from an historical perspective the
development of our understanding of how the sun (the nearest star)
shines, beginning in the following section with the nineteenth-century
controversy over the age of the sun.  In later sections, we shall see
how seemingly unrelated discoveries in fundamental physics led to a
theory of nuclear energy generation in stars that resolved the
controversy over the age of the sun and explained the origin of solar
radiation. In the section just before the summary, we shall discuss
how experiments that were designed to test the theory of nuclear
energy generation in stars revealed a new mystery, the Mystery of the
Missing Neutrinos.

\section{The Age of the Sun}

How old is the sun? How does the sun shine? These questions are
two sides of the same coin, as we shall see.

The rate at which the sun is radiating energy is easily
computed by using the measured rate at which energy reaches the earth's
surface and the distance between the earth and the sun.  The total
energy that the sun has radiated away over its lifetime is
approximately the product of the current rate at which energy is being
emitted, which is called the solar luminosity, times the age of the sun.

The older the sun is, the greater the total amount of radiated solar
energy. The greater the radiated energy, or the larger the age of the
sun, the more difficult it is to find an explanation of the source of
solar energy.

To better appreciate how difficult it is to find an explanation, let
us consider a specific illustration of the enormous rate at which the
sun radiates energy.  Suppose we put a cubic centimeter of ice outside
on a summer day in such a way that all of the sunshine is
absorbed by the ice.  Even at the great distance between the earth and
the sun, sunshine will melt the ice cube in about $40$ minutes.  Since
this would happen anywhere in space at the earth's distance from the
sun, a huge spherical shell of ice centered on the sun and 300 million
km ($200$ million miles) in diameter would be melted in the same
time.  Or, shrinking the same amount of ice down to the surface of the
sun, we can calculate that an area ten thousand times the area of the
earth's surface and about half a kilometer ($0.3$ mile) thick would
also be melted in $40$ minutes by the energy pouring out of the sun.

In this section, we shall discuss how nineteenth-century scientists
tried to determine  the source of solar energy, using the solar age as
a clue.

\subsection{Conflicting Estimates of the Solar Age}

The energy source for solar radiation was believed by
nineteenth-century physicists to be gravitation. In an influential
lecture in 1854, Hermann von Helmholtz, a German professor of
physiology who became a distinguished researcher and physics
professor, proposed that the origin of the sun's enormous radiated
energy is the gravitational contraction of a large mass.  Somewhat
earlier, in the 1840's, J. R. Mayer (another German physician) and
J. J. Waterson had also suggested that the origin of solar radiation
is the conversion of gravitational energy into
heat\footnote{\tightenlines von
Helmholtz and Mayer were two of  the codiscoverers of the law of
conservation of energy. This law states that energy can be transformed
from one form to another but the total amount is always
conserved. Conservation of energy is a basic principle of modern
physics that is used in analyzing the very smallest (sub-atomic)
domains and the largest known structure (the universe), and just about
everything in between. We shall see later that Einstein's
generalization of the law of conservation of energy was a key
ingredient in understanding the origin of solar radiation. The
application of conservation of energy to radioactivity revealed the
existence of neutrinos.}.

Biologists and geologists considered the effects of solar radiation,
while physicists concentrated on the origin of the radiated energy.
In 1859, Charles Darwin, in the first edition of {\it On The Origin of
the Species by Natural Selection}, made a crude calculation of the age
of the earth by estimating how long it would take erosion occurring at
the current observed rate to wash away the Weald, a great valley that
stretches between the North and South Downs across the south of
England. He obtained a number for the ``denudation of the Weald'' in
the range of $300$ million years, apparently long enough for natural
selection to have produced the astounding range of species that exist
on earth.

As Herschel stressed, the sun's heat is responsible for life
and for most geological evolution on earth. Hence, Darwin's estimate
of a minimum age for geological activity on the earth implied a
minimum estimate for the amount of energy that the sun has radiated.

Firmly opposed to Darwinian natural selection, William Thompson, later
Lord Kelvin, was a professor at the University of Glasgow and one of
the great physicists of the nineteenth century. In addition to his
many contributions to applied science and to engineering, Thompson
formulated the second law of thermodynamics and set up the absolute
temperature scale, which was subsequently named the Kelvin scale in
his honor.  The second law of thermodynamics states that heat
naturally flows from a hotter to a colder body, not the opposite.
Thompson therefore realized that the sun and the earth must get colder
unless there is an external energy source and that eventually the
earth will become too cold to support life.

Kelvin, like Helmholtz,  was convinced that the sun's luminosity was
produced by the conversion of gravitational energy into heat.  In an
early (1854) version of this idea, Kelvin suggested that the sun's
heat might be produced continually by the impact of meteors falling
onto its surface.  Kelvin was forced by astronomical evidence to modify
his hypothesis and he then argued that the primary source of the
energy available to the sun was the gravitational energy of the
primordial meteors from which it was formed.

Thus, with great authority and eloquence Lord Kelvin declared in 1862:

\begin{quote}
That some form of the meteoric theory is certainly the true and
complete explanation of solar heat can scarcely be doubted, when the
following reasons are considered: (1) No other natural explanation,
except by chemical action, can be conceived. (2) The chemical theory
is quite insufficient, because the most energetic chemical action we
know, taking place between substances amounting to the whole sun's
mass, would only generate about 3,000 years' heat. (3) There is no
difficulty in accounting for 20,000,000 years' heat by the meteoric
theory.
\end{quote}
Kelvin continued by attacking Darwin's estimate directly, asking
rhetorically: 
\begin{quote}
What then are we to think of such geological estimates
as [Darwin's] $300,000,000$ years for the ``denudation of the Weald''?
\end{quote}
Believing Darwin was wrong in his estimate of the age of the earth,
 Kelvin also believed that Darwin was wrong about the time available
 for natural selection to operate.

Lord Kelvin estimated the lifetime of the sun, and by implication the
earth, as follows. He calculated the gravitational energy of an object
with a mass equal to the sun's mass and a radius equal to the sun's
radius and divided the result  by the rate at which the sun radiates
away energy. This calculation yielded a lifetime of only $30$ million
years. The corresponding estimate for the lifetime sustainable by
chemical energy was much smaller because chemical processes release
very little energy.

\subsection{Who was right?}
\label{susbsec:whoright}

As we have just seen, in the nineteenth century you could get very
different estimates for the age of the sun, depending upon whom you
asked.  Prominent theoretical physicists argued, based upon the
sources of energy that were known at that time, that the sun was at
most a few tens of million years old.  Many geologists and biologists
concluded that the sun must have been shining for at least several
hundreds of millions of years in order to account for geological
changes and the evolution of living things, both of which depend
critically upon energy from the sun.  Thus the age of the sun, and the
origin of solar energy, were important questions not only for physics
and astronomy, but also for geology and biology.

Darwin was so shaken by the power of Kelvin's analysis and by the
authority of his theoretical expertise that in the last editions of
{\it On The Origin of the Species} he eliminated all mention of specific time
scales. He wrote in 1869 to Alfred Russel Wallace, the codiscoverer
of natural selection, complaining about Lord Kelvin:
\begin{quote}
Thompson's views on the recent age of the world have been for some
time one of my sorest troubles.
\end{quote}

Today we know that Lord Kelvin was wrong and the geologists and
evolutionary biologists were right. Radioactive dating of meteorites
shows that the sun is $4.6$ billion years old. 

What was wrong with Kelvin's analysis? An analogy may help.  Suppose a
friend observed you using your computer and tried to figure out how
long the computer had been operating. A plausible estimate might be no
more than a few hours, since that is the maximum length of time over
which a battery could supply the required amount of power. The flaw in
this analysis is the assumption that your computer is necessarily
powered by a battery. The estimate of a few hours could be wrong if
you computer were operated from an electrical power outlet in the
wall. The assumption that a battery supplies the power for your
computer is analogous to Lord Kelvin's assumption that gravitational
energy powers the sun.

Since nineteenth century theoretical physicists did not know about the
possibility of transforming nuclear mass into energy, they calculated
a maximum age for the sun that was too short. Nevertheless, Kelvin and
his colleagues made a lasting contribution to the sciences of
astronomy, geology, and biology by insisting on the principle that
valid inferences in all fields of research must be consistent with the
fundamental laws of physics.

We will now discuss some of the landmark developments in the
understanding of how nuclear mass is used as the fuel for stars.

\section{A Glimpse of the solution}

The turning point in the battle between theoretical physicists and
empirical geologists and biologists occurred in 1896.  In the course of
an experiment designed to study x-rays discovered the previous year by
Wilhelm Roentgen, Henri Becquerel stored some uranium-covered plates
in a desk drawer next to photographic plates wrapped in dark
paper. Because it was cloudy in Paris for a couple of days,
Becquerel was not able to ``energize'' his photographic plates by
exposing them to sunlight as he had intended. On developing the
photographic plates, he found to his surprise strong images of his
uranium crystals. He had discovered natural radioactivity, due to
nuclear transformations of  uranium.

The significance of Becquerel's discovery became apparent in 1903,
when Pierre Curie and his young assistant, Albert Laborde, announced
that radium salts constantly release heat.  The most extraordinary
aspect of this new discovery was that radium radiated heat without
cooling down to the temperature of its surroundings. The radiation
from radium revealed a previously unknown source of energy.  William
Wilson and George Darwin almost immediately proposed that
radioactivity might be the source of the sun's radiated energy.

The young prince of experimental physics, Ernest Rutherford, then a
professor of physics at McGill University in Montreal, discovered the
enormous energy released by alpha particle radiation from radioactive
substances. In 1904, he announced: 
\begin{quote}
The discovery of the radio-active
elements, which in their disintegration liberate enormous amounts of
energy, thus increases the possible limit of the duration of life on
this planet, and allows the time claimed by the geologist and
biologist for the process of evolution.
\end{quote}

The discovery of radioactivity opened up the possibility that nuclear
energy might be the origin of solar radiation. This development freed
theorists from relying in their calculations on gravitational energy.
However, subsequent astronomical observations showed that the sun does
not contain a lot of radioactive materials, but instead is mostly
hydrogen in gaseous form. Moreover, the rate at which radioactivity
delivers energy does not depend on the stellar temperature, while
observations of stars suggested that the amount of energy radiated by
a star does depend sensitively upon the star's interior
temperature. Something other than radioactivity is required to release
nuclear energy within a star.

In the next sections, we shall trace the steps that led to what we now
believe is the correct understanding of how stars shine.

\section{The Direction Established}

The next fundamental advance came once again from an unexpected
direction. In 1905, Albert Einstein derived his famous relation
between mass and energy, $E = mc^2$, as a consequence of the special
theory of relativity. Einstein's equation showed that a tiny amount of
mass could, in principle, be converted into a tremendous amount of
energy. His relation generalized and extended the nineteenth century
law of conservation of energy of von Helmholtz and Mayer to include
the conversion of mass into energy.

What was the connection between Einstein's equation and the energy
source of the sun? The answer was not obvious. Astronomers did their
part by defining the constraints that observations of stars imposed on
possible explanations of stellar energy generation. In 1919, Henry
Norris Russell, the leading theoretical astronomer in the United
States, summarized in a concise form the astronomical hints on the
nature of the stellar energy source. Russell stressed that the most
important clue was the high temperature in the interiors of stars.

F. W. Aston discovered in 1920 the key experimental element in the
puzzle.  He made precise measurements of the masses of many different
atoms, among them hydrogen and helium. Aston found that four hydrogen
nuclei were heavier than a helium nucleus. This was not the principal
goal of the experiments he performed, which were motivated in large
part by looking for isotopes of neon.

\begin{figure}[!h]
\centerline{\psfig{figure=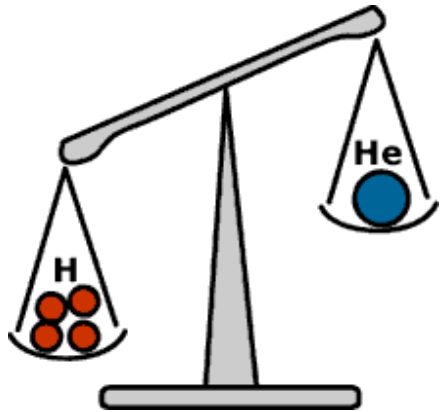,width=2.25in}}
\tightenlines
\caption[]{\footnotesize Aston showed in 1920 that four hydrogen nuclei are
heavier than a helium nucleus.
\label{fig:scale}}
\end{figure}

The importance of Aston's measurements was immediately recognized by
Sir Arthur Eddington, the brilliant English astrophysicist. Eddington
argued in his 1920 presidential address to the British Association for
the Advancement of Science that Aston's measurement of the mass
difference between hydrogen and helium meant that the sun could shine
by converting hydrogen atoms to helium.  This burning of hydrogen into
helium would (according to Einstein's relation between mass and
energy) release about $0.7$\% of the mass equivalent of the energy.
In principle, this could allow the sun to shine for about a 100
billion years.

In a frighteningly prescient insight, Eddington
went on to remark about the connection between stellar energy
generation and the future of humanity:
\begin{quote}
If, indeed, the sub-atomic energy in the stars is being freely used to
maintain their great furnaces, it seems to bring a little nearer to
fulfillment our dream of controlling this latent power for the
well-being of the human race---or for its suicide.
\end{quote}

\section{Understanding the Process}

The next major step in understanding how stars produce energy from
nuclear burning, resulted from applying quantum mechanics to the
explanation of nuclear radioactivity. This application was made
without any reference to what happens in stars.  According to
classical physics, two particles with the same sign of electrical
charge will repel each other, as if they were repulsed by a mutual
recognition of `bad breath'.  Classically, the probability that two
positively charged particles get very close together is zero.  But,
some things that cannot happen in classical physics can occur in the
real world which is described on a microscopic scale by quantum
mechanics.

In 1928, George Gamow, the great Russian-American theoretical
physicist, derived a quantum-mechanical formula that gave a non-zero
probability of two charged particles overcoming their mutual
electrostatic repulsion and coming very close together.  This quantum
mechanical probability is now universally known as the ``Gamow
factor.'' It is widely used to explain the measured rates of certain
radioactive decays.

In the decade that followed Gamow's epochal work, Atkinson and
Houtermans and later Gamow and Teller used the Gamow factor to derive
the rate at which nuclear reactions would proceed at the high
temperatures believed to exist in the interiors of stars.  The Gamow
factor was needed in order to estimate how often two nuclei with the
same sign of electrical charge would get close enough together to fuse
and thereby generate energy according to Einstein's relation between
excess mass and energy release.

In 1938, C. F. von Weizs\"acker came close to solving the problem of
how some stars shine. He discovered a nuclear cycle, now known as the
carbon-nitrogen-oxygen (CNO) cycle (see Appendix A), in which hydrogen nuclei could be
burned using carbon as a catalyst. However, von Weizs\"acker did not
investigate the rate at which energy would be produced in a star by
the CNO cycle nor did he study the crucial dependence upon stellar
temperature.

By April 1938, it was almost as if the scientific stage had been
intentionally set for the entry of Hans Bethe, the acknowledged master
of nuclear physics. Professor Bethe had just completed a classic set
of three papers in which he reviewed and analyzed all that was then
known about nuclear physics. These works were known among his
colleagues as
``Bethe's bible.'' Gamow assembled a small conference of physicists and
astrophysicists in Washington, D. C. to discuss the state of
knowledge, 
and the unsolved problems, concerning the internal constitution of the stars.

In the course of the next six months or so, Bethe worked out the basic
nuclear processes by which hydrogen is burned (fused) into helium in
stellar interiors.  Hydrogen is the most abundant constituent of the
sun and similar stars, and indeed the most abundant element in the
universe.

Bethe described the results of his calculations in a paper entitled
``Energy Production in Stars,'' which is awesome to read. He
authoritatively analyzed the different possibilities for reactions
that burn nuclei and selected as most important the two processes that
we now believe are responsible for sunshine. One process, the
so-called $p-p$ chain (see Appendix B), builds helium out of hydrogen and is the
dominant energy source in stars like the sun and less massive stars.

The CNO cycle, the second process which was also considered by von
Weizs\"acker, is most important in stars that are more massive than
the sun. Bethe used his results to estimate the central temperature of
the sun and obtained a value that is within $20$\% of what we
currently believe is the correct value ($16$ million degrees
Kelvin)\footnote{\tightenlines According to the modern theory of stellar evolution,
the sun is heated to the enormous temperatures at which nuclear fusion
can occur by gravitational energy released as the solar mass contracts
from an initially large gas cloud. Thus Kelvin and other
nineteenth-century physicists were partially right; the release of
gravitational energy ignited nuclear energy generation in the sun. }.
Moreover, he showed that his calculations led to a relation between
stellar mass and stellar luminosity that was in satisfactory agreement
with the available astronomical observations.

In the first two decades after the end of the second world war, many
important details were added to Bethe's theory of nuclear burning in
stars. Distinguished physicists and astrophysicists, especially
A. G. W. Cameron, W. A. Fowler, F. Hoyle, E. E. Salpeter,
M. Schwarzschild,  and their experimental colleagues, returned eagerly
to the question of how stars like the sun generate energy.  From
Bethe's work, the answer was known in principle: the sun produces the
energy it radiates by burning hydrogen. According to this theory, the
solar interior is a sort of controlled thermonuclear bomb on a giant
scale\footnote{\tightenlines The sensitive dependence of the Gamow factor upon the
relative energy of the two charged particles is, we now understand,
what provides the temperature ``thermostat'' for stars.}.  The theory
leads to the successful calculation of the observed luminosities of
stars similar to the sun and provides the basis for our current
understanding of how stars shine and evolve over time.  The idea that
nuclear fusion powers stars is one of the cornerstones of modern
astronomy and is used routinely by scientists in interpreting
observations of stars and galaxies.

W. A. Fowler,Willy  as he was universally known, led a team of colleagues in
his Caltech Kellogg Laboratory and inspired physicists throughout the
world to measure or calculate the most important details of the $p-p$
chain and the CNO cycle. There was plenty of work to do and the
experiments and the calculations were difficult. But, the work got
done because understanding the specifics of solar energy generation
was so interesting. Most of the efforts of Fowler and his
colleagues M. Burbidge, G. R. Burbidge, F. Hoyle, and
A. G. W. Cameron) soon shifted to the problem of how the heavy
elements, which are needed for life, are produced in stars.

\section{Testing the Hypothesis of Nuclear Burning}

Science progresses as a result of the clash between theory and
experiment, between speculation and measurement.  Eddington, in the
same lecture in which he first discussed the  burning of hydrogen nuclei in
stars, remarked:
\begin{quote}
I suppose that the applied mathematician whose theory has just passed
one still more stringent test by observation ought not to feel
satisfaction, but rather disappointment---``Foiled again! This time I {\it
had} hoped to find a discordance which would throw light on the points
where my model could be improved.'' 
\end{quote}

Is there any way to test the theory that the sun shines because very
deep in its interior hydrogen is burned into helium? At first thought,
it would seem impossible to make a direct test of the nuclear burning
hypothesis. Light takes about ten million years to leak out from the
center of the sun to the surface and when it finally emerges in the
outermost regions, light mainly tells us about the conditions in those
outer regions.  Nevertheless, there is a way of ``seeing'' into the
solar interior with neutrinos, exotic particles discovered while
trying to understand a different mystery\footnote{\tightenlines The existence of
neutrinos was first proposed by Wolfgang Pauli in a 1930 letter to his
physics colleagues as a "desperate way out" of the apparent
non-conservation of energy in certain radioactive decays (called
$\beta$-decays) in which electrons were emitted. According to Pauli's
hypothesis, which he put forward very hesitantly, neutrinos are
elusive particles which escape with the missing energy in
$\beta$-decays. The mathematical theory of $\beta$-decay was
formulated by Enrico Fermi in 1934 in a paper which was rejected by
the journal {\it Nature} because ``it contained speculations too
remote from reality to be of interest to the reader.'' Neutrinos from
a nuclear reactor were first detected by Clyde Cowan and Fred Reines
in 1956.}.

\subsection{Discovery, Confirmation, and Surprise}
\label{subsec:confirm}

A neutrino is a sub-atomic particle that interacts weakly with matter
and travels at a speed that is essentially the speed of light.
Neutrinos are produced in stars when hydrogen nuclei are burned to
helium nuclei; neutrinos are also produced on earth in particle
accelerators, in nuclear reactors, and in natural radioactivity.
Based upon the work of Hans Bethe and his colleagues, we believe that
the process by which stars like the sun generate energy can be
symbolized by the relation,

\begin{equation}
4 \hbox{$^{1}$H} \longrightarrow \hbox{$^{4}$He} ~+~2 e^+ ~+~ 2 
\hbox{$\nu_e$} ~+~{\rm energy},
\label{eqn:fusion}
\end{equation}
in which four hydrogen nuclei ($^1$H, protons) are burned into a
single helium nucleus ($^4$He, $\alpha$ particle) plus two positive
electrons ($e^+$) and two neutrinos ($\nu_e$) plus energy.  This
process releases energy to the star since, as Aston showed, four
hydrogen atoms are heavier than one helium atom.  The same set of
nuclear reactions that supply the energy of the sun's radiation also
produce neutrinos that can be searched for in the laboratory.

\begin{figure}[!t]
\centerline{\psfig{figure=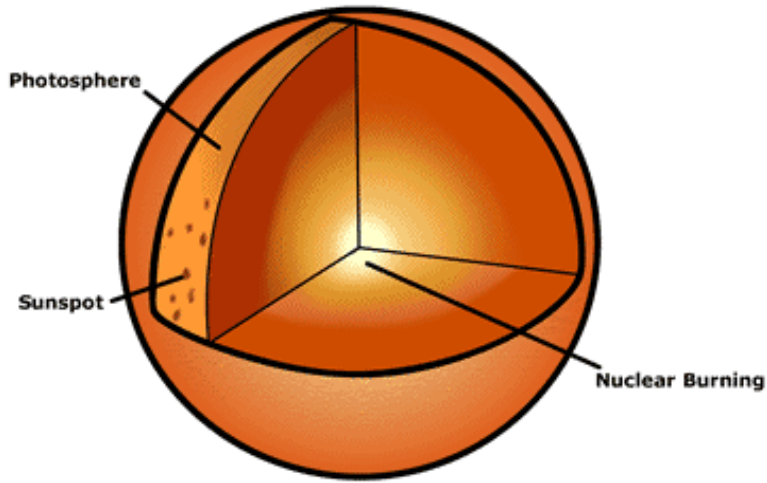,width=4in}}
\tightenlines
\vglue.2in
\caption[]{\footnotesize This figure is a cross section of the sun. The features
that are usually studied by astronomers with normal telescopes
that detect light are labeled on the outside, e.g., sunspot and prominences.
Neutrinos enable us to look deep inside the sun, into the solar core where
nuclear burning occurs.
\label{fig:insidesun}}
\end{figure}

Because of their weak interactions, neutrinos are difficult to detect.
How difficult?  A solar neutrino passing through the entire earth has
less than one chance in a thousand billion of being stopped by
terrestrial matter. According to standard theory, about a hundred
billion solar neutrinos pass through your thumbnail every second and
you don't notice them.  Neutrinos can travel unaffected through iron
as far as light can travel in a hundred years through empty space.

In 1964, Raymond Davis Jr. and I proposed that an experiment with
$100,000$ gallons of cleaning fluid (perchloroethylene, which is
mostly composed of chlorine) could provide a critical test of the idea
that nuclear fusion reactions are the ultimate source of solar
radiation. We argued that, if our understanding of nuclear processes
in the interior of the sun was correct, then solar neutrinos would be
captured at a rate Davis could measure with a large tank filled with
cleaning fluid.  When neutrinos interact with chlorine, they
occasionally produce a radioactive isotope of argon. Davis had shown
previously that he could extract tiny amounts of neutrino-produced
argon from large quantities of perchloroethylene. To do the solar
neutrino experiment, he had to be spectacularly clever since according
to my calculations, only a few atoms would be produced per week in a
huge Olympic-sized swimming pool of cleaning fluid.

Our sole motivation for urging this experiment was to
use neutrinos to:
\begin{quote}
enable us to see into the interior of a star and thus verify
directly the hypothesis of nuclear energy generation in stars.
\end{quote} 
As we shall see, Davis and I did not anticipate some of the most
interesting aspects of this proposal.

Davis performed the experiment and in 1968 announced the first
results. He measured fewer neutrinos than I predicted. As the
experiment and the theory were refined, the disagreement appeared more
robust.  Scientists rejoiced that solar neutrinos were detected but
worried why there were fewer neutrinos than predicted.

What was wrong? Was our understanding of how the sun shines incorrect?
Had I made an error in calculating the rate at which solar neutrinos
would be captured in Davis's tank? Was the experiment wrong? Or, did
something happen to the neutrinos after they were created in the sun?

Over the next twenty years, many different possibilities were examined
by hundreds, and perhaps thousands, of physicists, chemists, and
astronomers\footnote{\tightenlines Perhaps the most imaginative proposal was made by
Stephen Hawking, who suggested that the central region of the sun
might contain a small black hole and that this could be the reason why
the number of neutrinos observed is less than the predicted number.}.
Both the experiment and the theoretical calculation appeared to be
correct.

Once again experiment rescued pure thought. In 1986, Japanese
physicists led by Masatoshi Koshiba and Yoji Totsuka, together with
their American colleagues, Eugene Beier and Alfred Mann,
reinstrumented a huge tank of water designed to measure the stability
of matter. The experimentalists increased the sensitivity of their
detector so that it could also serve as a large underground
observatory of solar neutrinos.  Their goal was to explore the reason
for the quantitative disagreement between the predicted and the
measured rates in the chlorine experiment.

The new experiment (called Kamiokande) in the Japanese Alps also
detected solar neutrinos.  Moreover, the Kamiokande experiment
confirmed that the neutrino rate was less than predicted by standard
physics and standard models of the sun and demonstrated that the
detected neutrinos came from the sun.  Subsequently, experiments in
Russia (called SAGE, led by V. Gavrin), in Italy (GALLEX and later GNO
led by T. Kirsten and E. Belotti, respectively), and again in Japan
(Super-Kamiokande, led by Y. Totsuka and Y. Suzuki), each with
different characteristics, all observed neutrinos from the
solar interior. In each detector, the number of neutrinos observed was
somewhat lower than  standard theory predicted.

What do all of these experimental results mean?  

Neutrinos produced in the center of the sun have been detected in five
experiments. Their detection shows directly that the source of the
energy that the sun radiates is the fusion of hydrogen nuclei in the
solar interior. The nineteenth century debate between theoretical
physicists, geologists, and biologists has been settled empirically.

From an astrophysical perspective, the agreement between neutrino
observations and theory is  good. The observed energies of
the solar neutrinos match the values predicted by theory.  The rates
at which neutrinos are detected are less than predicted but not by a
large factor.  The predicted neutrino arrival rate at the earth
depends approximately upon the 25th power of the central temperature
of the sun, $T\times T\times ...T$ (25 factors of the temperature
$T$).  The agreement that has been achieved (agreement within a factor
of three) shows that we have empirically measured the central
temperature of the sun to an accuracy of a few percent. Incidentally,
if someone had told me in 1964 that number of neutrinos observed from the
sun would be within a factor of three of the predicted value, I would
have been astonished and delighted.

In fact, the agreement between normal astronomical observations (using
light rather than neutrinos) and theoretical calculations of solar
characteristics is much more precise.  Studies of the internal
structure of the sun using the solar equivalent of terrestrial
seismology (i. e., observations of solar vibrations) show that the
predictions of the standard solar model for the temperatures in the
central regions of the sun are consistent with the observations to an
accuracy of at least $0.1$\%.  In this standard model, the current age
of the sun is five billion years, which is consistent with the minimum
estimate of the sun's age made by nineteenth-century geologists and
biologists (a few hundred million years).

Given that the theoretical models of the sun describe astronomical
observations accurately, what can explain the disagreement by a factor
of two or three between the measured and the predicted solar neutrino
rates?

\subsection{New physics}
\label{subsec:newphysics}

Physicists and astronomers were once again forced to reexamine their
theories. This time, the discrepancy was not between different
estimates of the sun's age, but rather between predictions based upon
a widely accepted theory and direct measurements of particles produced
by nuclear burning in the sun's interior. This situation was sometimes
referred to as the Mystery of the Missing Neutrinos or, in language
that sounded more scientific, the Solar Neutrino Problem.

As early as 1969, two scientists working in Russia, Bruno Pontecorvo
and Vladimir Gribov, proposed that the discrepancy between standard
theory and the first solar neutrino experiment could be due to an
inadequacy in the textbook description of particle physics, rather
than in the standard solar model. (Incidentally, Pontecorvo was the
first person to propose using a chlorine detector to study neutrinos.)
Gribov and Pontecorvo suggested that neutrinos suffer from a multiple
personality disorder, that they oscillate back and forth between
different states or types.  

According to the suggestion of Gribov and Pontecorvo, neutrinos are
produced in the sun in a mixture of individual states, a sort of split
personality.  The individual states have different, small masses,
rather than the zero masses attributed to them by standard particle
theory. As they travel to the earth from the sun, neutrinos oscillate
between the easier-to-detect neutrino state and the more
difficult-to-detect neutrino state.  The chlorine experiment only
detects neutrinos in the easier-to-observe state.  If many of the
neutrinos arrive at earth in the state that is difficult to observe,
then they are not counted. It is as if some or many of the neutrinos
have vanished, which can explain the apparent mystery of the missing
neutrinos.

Building upon this idea, Lincoln Wolfenstein in 1978 and Stanislav
Mikheyev and Alexei Smirnov in 1985 showed that the effects of matter
on neutrinos moving through the sun might increase the oscillation
probability of the neutrinos if Nature has chosen to give them masses
in a particular range.

Neutrinos are also produced by the collisions of cosmic ray particles
with other particles in the earth's atmosphere. In 1998, the
Super-Kamiokande team of experimentalists announced that they had
observed oscillations among atmospheric neutrinos.  This finding
provided indirect support for the theoretical suggestion that solar
neutrinos oscillate among different states. Many scientists working in
the field of solar neutrinos believe that, in retrospect, we have had
evidence for oscillations of solar neutrinos since 1968.

But, we do not yet know what causes the multiple personality disorder
of solar neutrinos. The answer to this question may provide a clue to
physics beyond the current standard models of sub-atomic
particles. Does the change of identity occur while the neutrinos are
traveling to the earth from the sun, as originally proposed by Gribov
and Pontecorvo?  Or does matter cause solar neutrinos to ``flip out''?
Experiments are underway in Canada, Italy (three experiments), Japan
(two experiments), Russia, and the United States that are attempting
to determine the cause of the oscillations of solar neutrinos, by
finding out how much they weigh and how they transform from one type
to another. Non-zero neutrino masses may provide a clue to a still
undiscovered realm of physical theory.

\section{Nature: a wonderful mystery}

Nature has written a wonderful mystery. The plot continually changes
and the most important clues come from seemingly unrelated
investigations. These sudden and drastic changes of scientific scene
appear to be Nature's way of revealing the unity of all fundamental science.

The mystery begins in the middle of the nineteenth century with the
puzzle: How does the sun shine?  Almost immediately, the plot switches
to questions about how fast natural selection occurs and at what rate
geological formations are created. The best theoretical physics of the
nineteenth century gave the wrong answer to all these questions. The
first hint of the correct answer came, at the very end of the
nineteenth century, from the discovery of radioactivity with
accidentally darkened photographic plates. 

The right direction in which to search for the detailed solution was
revealed by the 1905 discovery of the special theory of relativity, by
the 1920 measurement of the nuclear masses of hydrogen and helium, and
by the 1928 quantum mechanical explanation of how charged particles
get close to each other. These crucial investigations were not
directly related to the study of stars.

By the middle of the twentieth century, nuclear physicists
and astrophysicists could calculate theoretically the rate of nuclear
burning in the interiors of stars like the sun. But, just when we
thought we had Nature figured out, experiments showed that fewer solar
neutrinos were observed at earth than were predicted by the standard
theory of how stars shine and how sub-atomic particles behave. 

At the beginning of the twenty-first century, we have learned that
solar neutrinos tell us not only about the interior of the sun, but
also something about the nature of neutrinos. No one knows what
surprises will be revealed by the new solar neutrino experiments that
are currently underway or are planned. The richness and the humor with
which Nature has written her mystery, in an international language
that can be read by curious people of all nations, is beautiful,
awesome, and humbling.

\vfill\eject

\appendix

\section{The CNO cycle}
\centering
\begin{minipage}{4in}
\centerline{\psfig{figure=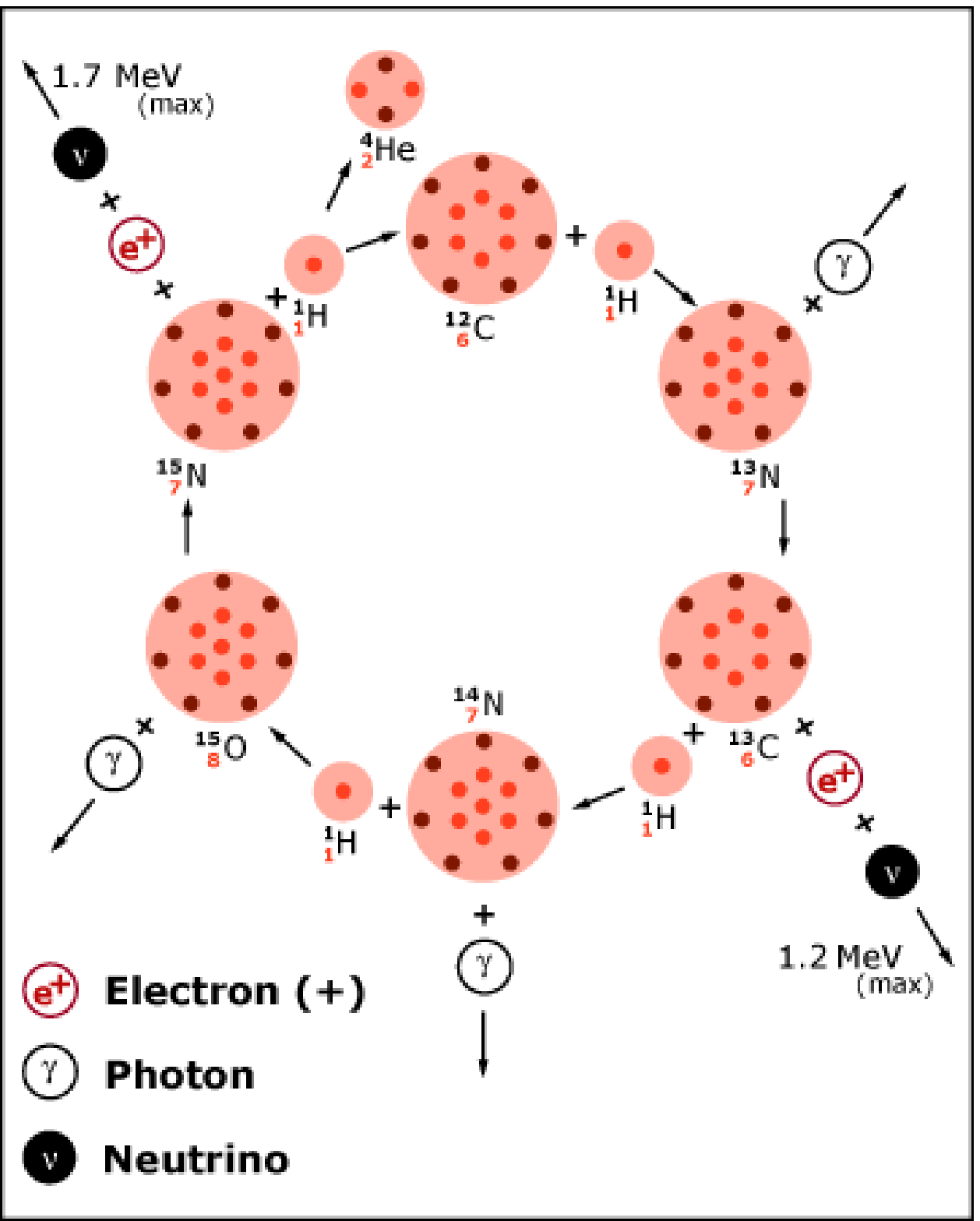,width=4in}}
\tightenlines\footnotesize
\vglue.2in
\noindent
For stars heavier than the sun, theoretical models
show that the CNO (carbon-nitrogen-oxygen) cycle of nuclear fusion is
the dominant source of energy generation. The cycle results in the
fusion of four hydrogen nuclei ($^1$H, protons) into a single helium
nucleus ($^4$He, alpha particle), which  supplies energy to the
star in accordance with Einstein's equation.  Ordinary carbon, $^{12}$C,
serves as a catalyst in this set of reactions and is regenerated. Only
relatively low energy neutrinos ($\nu$)are produced in this cycle. The
figure is adapted from J. N. Bahcall, {\it Neutrinos from the Sun},
Scientific American, Volume 221, Number 1, July 1969, pp. 28-37.
\end{minipage}
\vfill\eject
\section{The $p-p$ chain reaction}
\noindent
\centering
\vglue.15in
\begin{figure}[!h]
\centerline{\psfig{figure=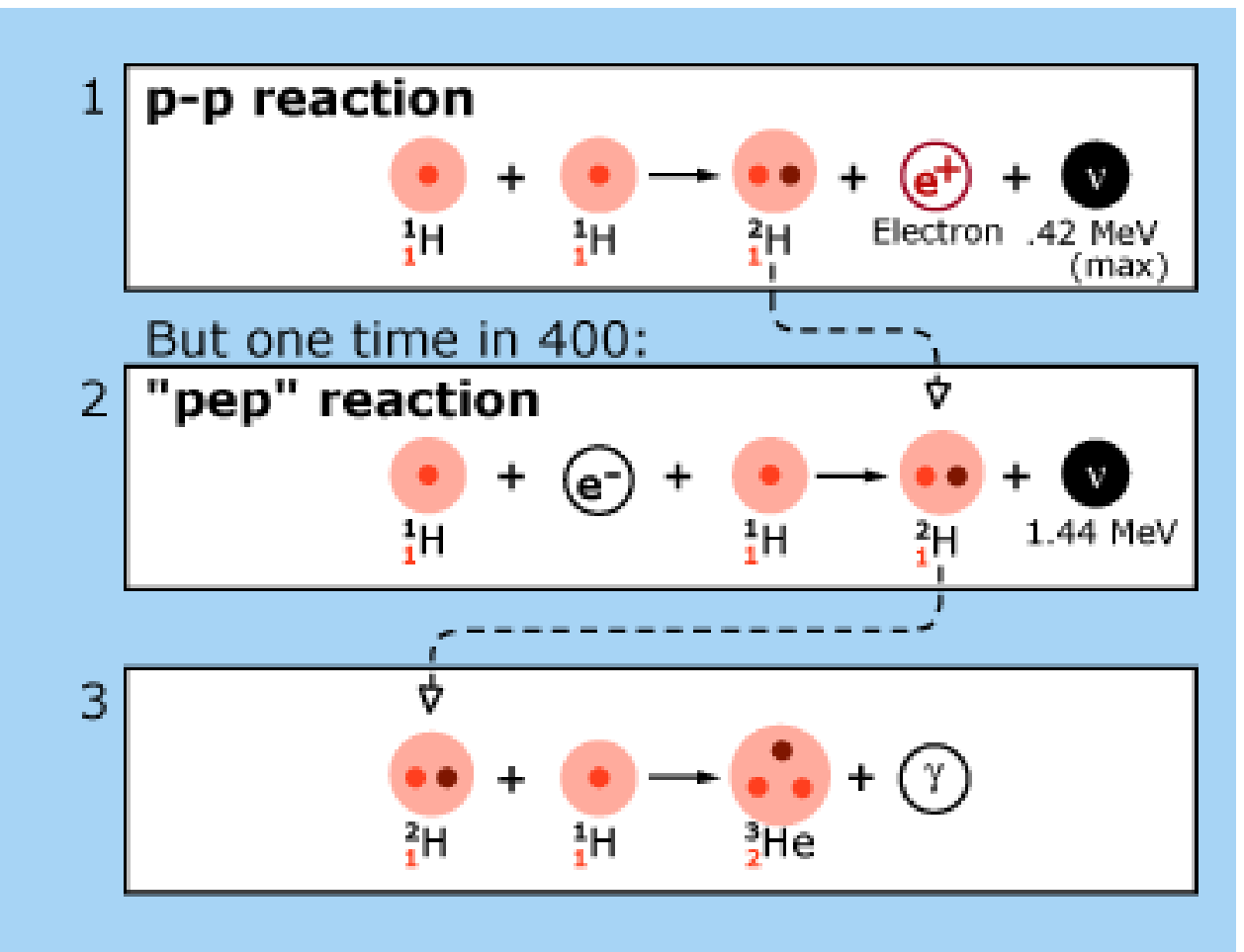,width=3.7in}}
\vglue.2in
\centerline{\psfig{figure=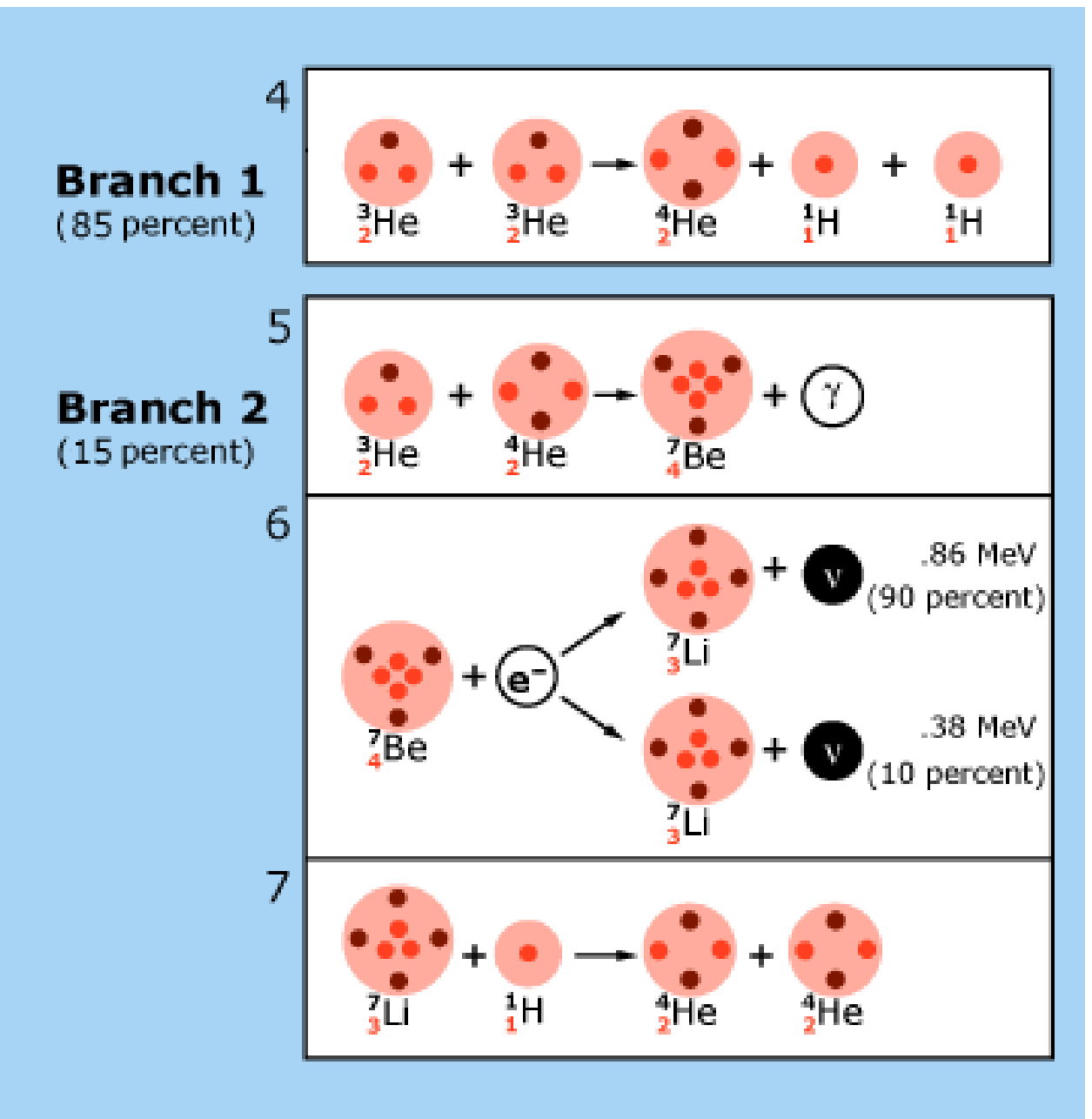,width=3.7in}}
\end{figure}

\centering
\begin{minipage}{3.7in}
\centerline{\psfig{figure=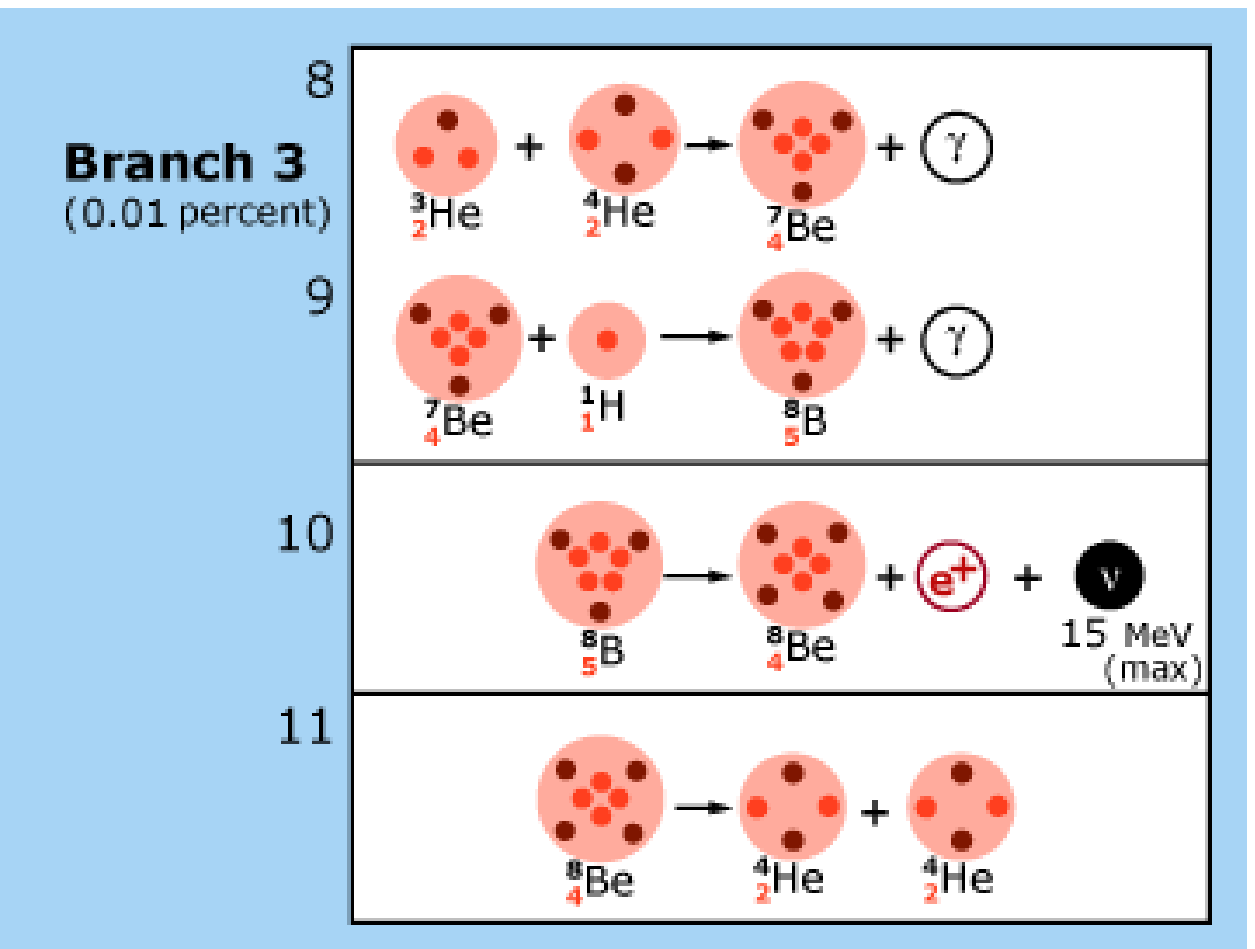,width=3.7in}}
\tightenlines\footnotesize
\vglue.15in
\noindent
In theoretical models of the sun, the $p-p$ chain of
nuclear reactions illustrated here is the dominant source of energy
production. Each reaction is labeled by a number in the upper left
hand corner of the box in which it is contained.  In reaction 1, two
hydrogen nuclei ($^1$H, protons) are fused to produced a heavy
hydrogen nucleus ($^2$H, a deuteron).  This is the usual way nuclear
burning gets started in the sun. On rare occasions, the process is
started by reaction 2.  Deuterons produced in reactions 1 and 2 fuse
with protons to produce a light element of helium ($^3$He). At this
point, the $p-p$ chain breaks into three branches, whose relative
frequencies are indicated in the figure.  The net result of this chain
is the fusion of four protons into a single ordinary helium nucleus
($^4$He) with energy being released to the star in accordance with
Einstein's equation.  Particles called `neutrinos' ($\nu$) are emitted
in these fusion processes. Their energies are shown in the figure in
units of millions of electron volts (MeV).  Reactions 2 and 4 were not
discussed by Hans Bethe.  This figure is adapted from J. N. Bahcall,
{\it Neutrinos from the Sun}, Scientific American, Volume 221, Number
1, July 1969, pp. 28-37.
\end{minipage}

\end{document}